\documentclass[]{agujournal}
\draftfalse
\usepackage{hyperref}

\begin{document}

\title{Titan's variable ionosphere during the T118-T119 Cassini flybys}

\authors{N. J. T. Edberg\affil{1}, E. Vigren\affil{1}, D. Snowden\affil{2}, L. H. Regoli\affil{3}, O. Shebanits\affil{1,2}, J.-E. Wahlund\affil{1}, D. J. Andrews\affil{1}, C. Bertucci\affil{4}, J. Cui\affil{5,6}}

\affiliation{1}{Swedish Institute of Space Physics, L\"agerhyddsv\"agen 1, 75121, Uppsala, Sweden}
\affiliation{2}{Department of Physics, Imperial College London, UK}
\affiliation{3}{Department of Physics, Central Washington University, 400 E. University Way, Ellensburg, WA, 98926, USA}
\affiliation{4}{Climate and Space Sciences and Engineering, University of Michigan, Ann Arbor, USA}
\affiliation{5}{IAFE, Ciudad Universitaria, Buenos Aires, Argentina}
\affiliation{6}{School of Atmospheric Sciences, Sun Yat-sen University, Zhuhai, Guangdong 519082, China}
\affiliation{7}{Key Laboratory of Lunar and Deep Space Exploration, Chinese Academy of Sciences, Beijing 100012, China}

\correspondingauthor{N. J. T. Edberg}{ne@irfu.se}



\begin{keypoints}
\item Large difference in electron density between T118 and T119 despite similar flybys
\item The lowest density ever observed in Titan's deep nightside ionosphere  during T118
\item Low fluxes of impact ionising particles and ionospheric dynamics is suggested as the most probable explanation
\end{keypoints}

%
%


\begin{abstract}
A significant difference in Titan's ionospheric electron density is observed between the T118 and T119 Cassini nightside flybys. These flybys had similar geometry, occurred at the same Saturn local time and while Titan was exposed to similar EUV and ambient magnetic field conditions. Despite these similarities, the RPWS/LP measured density differed a factor of 5 between the passes. This difference was present, and similar, both inbound and outbound. Two distinct electron peaks were present during T118, at 1150 km and 1200 km, suggesting very localised plasma production. During T118, from 1200-1350 km and below 1100 km, the lowest electron density ever observed in Titan's ionosphere are reported. We suggest that an exceptionally low rate of particle impact ionisation in combination with increased dynamics in the ionosphere could be the cause. This is, however, not verified by measurements and the measured ambient high energy particle pressure is in fact higher during T118 than during T119.
\end{abstract}

%
%

 \section{Introduction}
Titan, the largest moon of Saturn, is surrounded by a dense and extended ionosphere \citep{bird1997, wahlund2005}. The structure and variability of this ionosphere are sensitive to several factors. On the dayside of the moon, the ionosphere is created mainly through photo-ionisation by Extreme Ultraviolet (EUV) radiation on the atmospheric N$_2$ and CH$_4$ \citep{agren2009,galand2010}, which means that Titan's ionospheric properties will vary with the solar rotation and the phase of the solar cycle \citep{edberg2013a,shebanits2017}. Particle impact ionisation also contributes, and is the main ionisation source on the nightside of the moon \citep{agren2007, cravens2008, vigren2013, vigren2015b}. The ambient plasma conditions in Saturn's magnetosphere, and in which region of Saturn's magnetosphere Titan is located in, can therefore influence the ionospheric structure of Titan significantly \citep{edberg2015b}. Another factor that is  important is the topology of the ambient magnetic field, which affects the particle precipitation into Titan's ionosphere \citep{regoli2016}. Diurnal variations and plasma transport also modulates especially the nightside ionosphere \citep{cui2009, cui2010, snowden2013}. 

The Cassini spacecraft sampled the ionosphere of Titan during 127 targeted flybys over a time period of 13 years. The flyby geometry was quite diverse around Titan and they occurred at different Saturn Local Times (SLT). The ambient plasma and magnetic field properties as well as the solar EUV radiation covered a large range of values over this period, which enabled studies of Titan under various conditions. Furthermore, some flybys were designed to occur in sequences, such that the flyby geometry gradually shifted from one pass to the next. The flyby pair T118 and T119 is one such example, and in this paper we will mainly focus on these two passes to compare and study the ionospheric structure and variability.

\subsection{Instruments and coordinate systems}
We use measurements from the Radio and Plasma Waves System/Langmuir Probe (RPWS/LP) instrument \citep{gurnett2004,wahlund2005} to obtain the electron density in Titan's ionosphere. We derive the electron density from the sweep mode measurements in the same way as described in e.g. \citet{edberg2011,shebanits2013}, and references therein. These density data are compared with the density obtained from the RPWS observations of the upper hybrid frequency ($F_{UH}$) emission line, and they agree to within a few percentage in the upper part of the ionosphere, while in the deeper part of the ionosphere there typically is a larger difference, up to a hundred cm$^{-3}$. It should be noted that this discrepancy has existed throughout the mission and is  not unique for these two flybys. 

Furthermore, we use magnetic field vector measurements from the Cassini magnetometer instrument (MAG) \citep{dougherty2004} at a cadence of 1 min and we use energetic particle measurements for H$+$ and W$+$ (W$^+$ indicates water group ions) from the Magnetosphere Imaging Instrument/Charge Energy Mass Spectrometer (MIMI/CHEMS) to obtain partial particle pressure at Titan's location during the flybys of interest. CHEMS can measure ions with energies from 2.8 keV to 220 keV for H$^+$ and from 8.9 keV to 220 keV for W$^+$, but not the lower energy range. There were no measurement of the lower energy plasma (few eV - $\sim$1 keV) such that we cannot assess the influence of the suprathermal plasma. However, on average, in the inner magnetosphere of Saturn, 50\% of the plasma pressure is carried by the energetic particles, at least out to 10 R$_S$ \citep{sergis2009}.

The two coordinate systems used in this paper are the Titan interaction (TIIS) coordinate system and the ecliptic coordinate system, see e.g. \citet{edberg2015b} for a description. 

\section{Observations}
Below we will first describe the specific flyby geometry during T118 and T119 before presenting the measurements from these two passes.

\subsection{The T118 and T119 flybys}
\begin{figure}[ht]
 \centering
 \includegraphics[width=25pc]{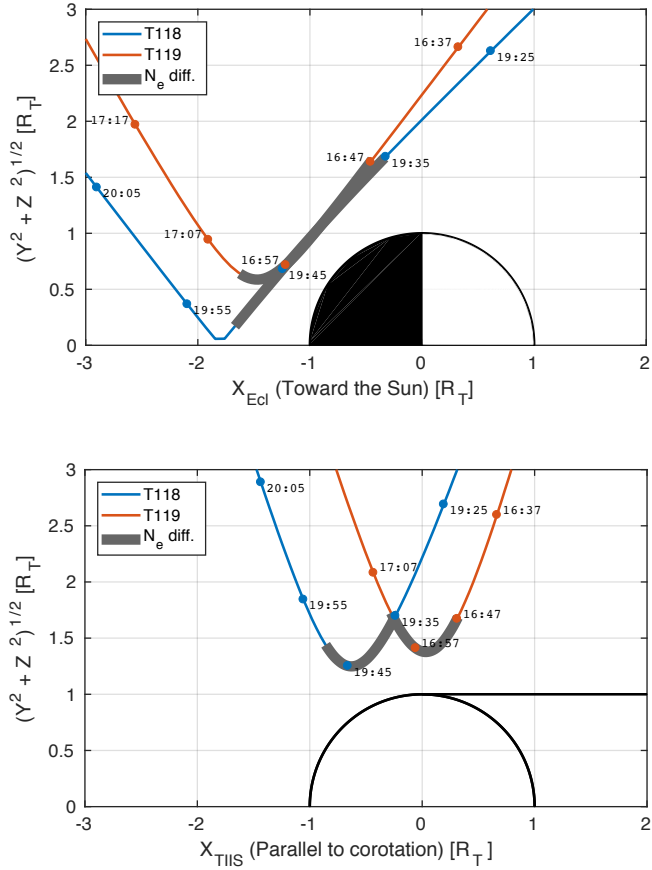}
 \caption{Flyby geometry during T118 and T119 in cylindrical (top) TIIS and (bottom) ecliptic coordinates. The thick gray line indicate the time when a significant difference in electron density between the two passes is present.}
 \label{figone}
\end{figure}

The flyby geometries during T118 and T119 were quite similar. During both the T118 and T119 flyby, Titan was located at 02h SLT and the passes had their C/A on the southern nightside of Titan, as can be seen in Figure \ref{figone}, and on the Saturn-averted side of Titan. 
Cassini approached Titan and had its inbound legs starting from the dayside of Titan, which in this case partly coincides with the magnetospheric plasma flow wake side of the moon. The flybys took place roughly one month apart, on 4 April 2016 and on 6 May 2016, respectively. There is a shift of the T119 C/A toward the wake side compared to T118, but the ionospheric structure is on average not varying much on such short spatial scales in the ram-wake direction \citep{edberg2015b}, such that this small difference is negligible for this study. Due to that part of the ionisation occurs through EUV radiation, any difference in solar zenith angle (SZA) could have some influence on the ionospheric structure, but in this case the SZA profiles does not differ much between the two flybys, as can be viewed in the upper panel of Figure \ref{figone} (the SZA variation is also shown as time series in Figure \ref{figtwo}e below). One difference that exists in between the two flybys is the attitude of Cassini (not shown). During T118, the LP was in the ram direction, i.e. pointing in the direction of the spacecraft motion, while during T119 it was turned by about 80$^{\circ}$. This should, however, not have any influence on the density measurements. The thick gray bands overlayed on the trajectory lines indicate where we observe very large differences in electron density between the two flybys, which will be described next.

\subsection{Measurements}
\begin{figure}[ht]
 \centering
 \includegraphics[width=40pc]{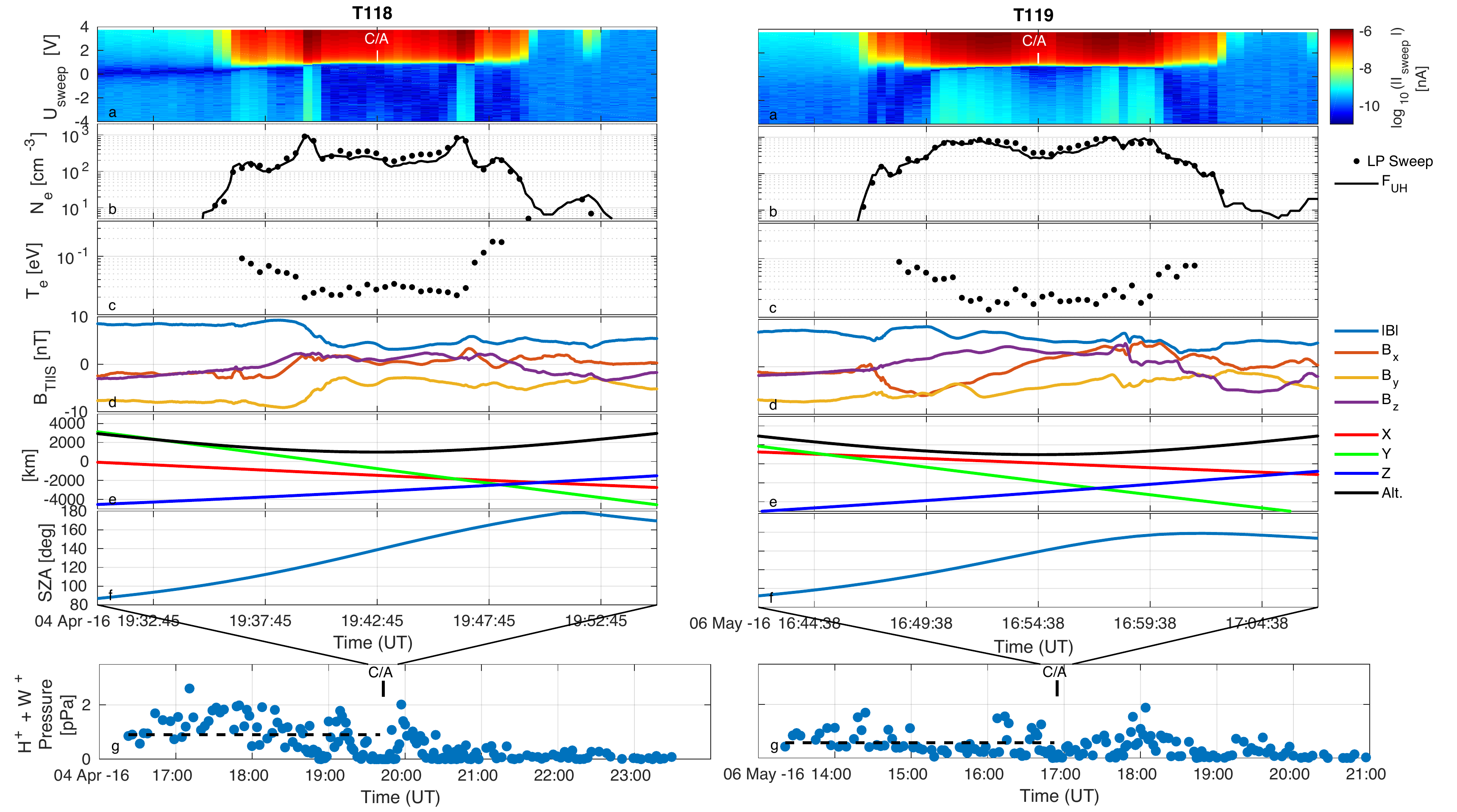}
\caption{Time series of Cassini data from the (left) T118 flyby and the (right) T119 flyby, centered around C/A. The panels show (a) LP sweeps, (b) electron density derived from the sweeps together with the electron density estimated from the $F_{UH}$, (c) electron temperature from the sweeps and restricted to when the electron density is above 100 cm$^{-3}$ (i.e. when the measurements are more reliable), (d) magnetic field magnitude and components in the TIIS coordinate system, (e) Cassini's position vector in TIIS as well as altitude, (f) SZA of Cassini, and (g) in a longer time interval, the ambient high energy particle pressure as measured by MIMI/CHEMS. The dashed lines indicate the mean pressure during the inbound legs.}
 \label{figtwo}
\end{figure}

Figure \ref{figtwo} presents the data from measurements taken during the T118 and T119 flybys, centered on the time of C/A. The top six panels (Figure \ref{figtwo}a-f) cover $\sim$25 min each while the lower single panels (Figure \ref{figtwo}g) cover 8h. The top panel shows the RPWS/LP sweep measurements, from which the electron density $N_e$ is derived. The electron density from the sweeps is shown in Figure \ref{figtwo}b together with the electron density estimate from the $F_{UH}$ line (proportional to electron density as $F_{UH} = 8980\sqrt{N_e}$) as detected by the RPWS antennas. 
The two density estimates agree very well in the topside ionosphere for both passes, but less so in the deepest part of the ionosphere. The electron temperature is quite similar in between the two flybys (Figure \ref{figtwo}c) and the magnetic field data (Figure \ref{figtwo}d) indicates very similar ambient magnetic field directions both before and after the flybys. The bottom panels show energetic particle pressure of H$^+$ + W$^+$ from the MIMI instrument during a longer interval in order to capture the ambient conditions prior to and after the flybys.

What should be particularly noted in Figure \ref{figtwo}, and which is central for this paper, is the large difference in electron density in between the two flybys in the interval of about $\pm$ 7 min around C/A $\sim$1000-1800 km). This can be seen when comparing the density estimates (Figure \ref{figtwo}b)  for the two flybys (noting again that there is also a difference between density derived from the LP and the density from the $F_{UH}$). The difference between T118 and T119 can also be clearly seen in the `raw' LP sweep data from Figure \ref{figtwo}a, where there is a striking difference in both collected ion and electron current (at positive and negative bias voltage values U$_{sweep}$, respectively) from one flyby to the next.
\begin{figure}[ht]
 \centering
 \includegraphics[width=35pc]{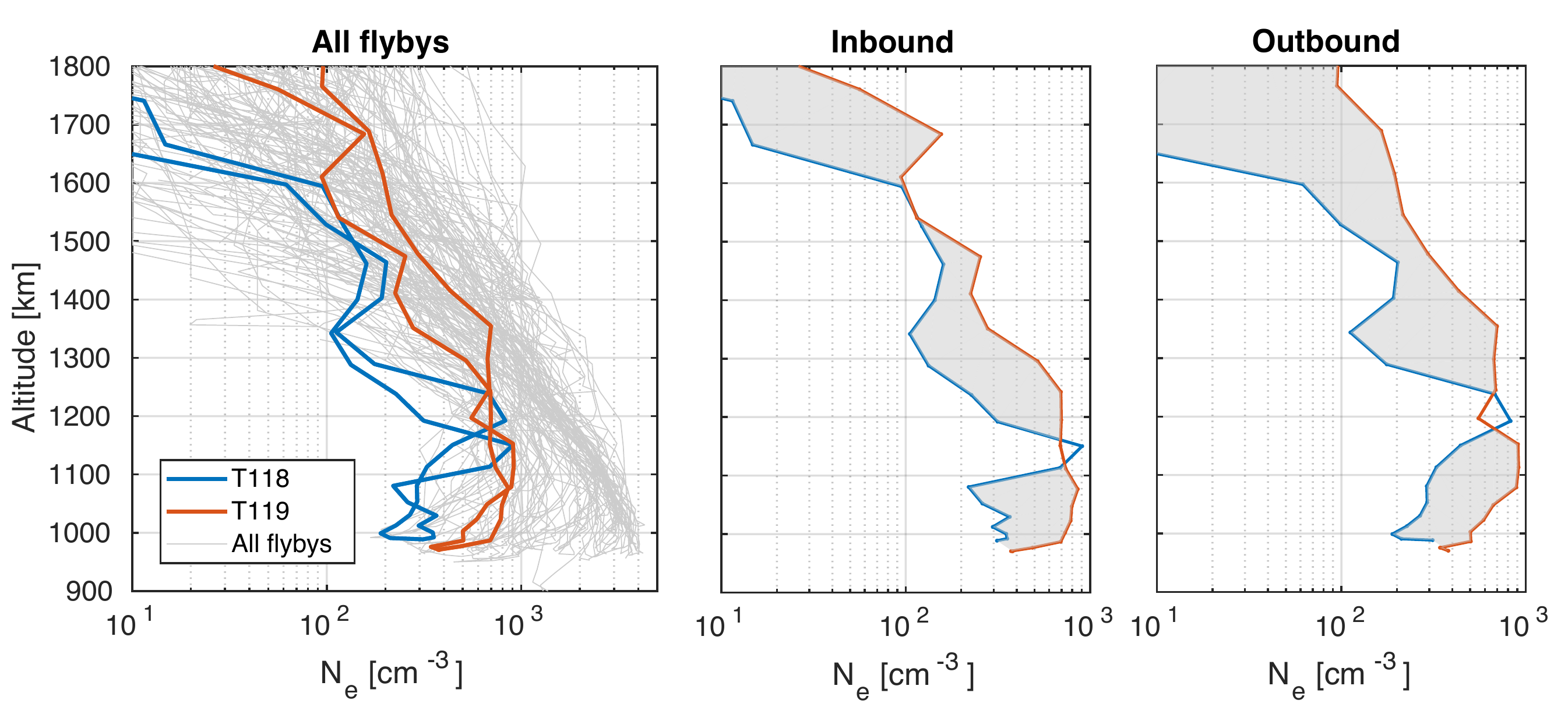}
 \caption{(Left) Altitude profiles of the electron density in Titan's ionosphere as measured by RPWS/LP during the T118 (blue) and the T119 (red) flybys. The gray thin lines show the electron density altitude profiles from all of Cassini's passes through Titan's ionosphere. (Middle and right) The same altitude profiles again but separated into inbound and outbound leg. The gray shaded area indicate the region of large density difference between the two passes.} 
 \label{figthree}
\end{figure}

To make this density difference more apparent we plot in Figure \ref{figthree} the altitude profiles of the LP measured electron density for the T118 (blue line) and the T119 (red line) flybys. From this plot it is evident that there is a significant difference in ionospheric density from the two flybys. The difference exceeds a factor of 5 at certain altitudes, e.g. around 1300-1400 km during the outbound leg, and is typically around a factor of 2-4 throughout  most of the interval. Also shown, with gray lines, are the electron density altitude profiles obtained from all of the other Cassini passes through Titan's ionosphere, for comparison. Looking at all of these profiles, there is a natural large variability in the ionospheric density spanning more than an order of magnitude for a given altitude. However, all of these profiles are gathered at a wide range of positions around Titan, at different SLT, at different phases of the solar cycle and during different ambient plasma conditions, which all affect the ionospheric structure of Titan \citep{edberg2015b}. What is intriguing during T118 and T119, is that all of these factors are more or less the same during the two passes: the SLT is 02h, the ambient conditions in Saturn's magnetosphere were typical of the northern magnetospheric lobe, although with stronger magnetic field fluctuations during T118, according to MAG data presented by \citet{kabanovic2017}, the EUV levels  as observed by the Solar Dynamic Observatory/Extreme Ultraviolet Variability Experiment (SDO/EVE) were similar and no large flare were observed from space observatories. 
During T118, it is striking that the measured electron density is very low for large parts of the ionosphere compared to all other profiles. For some altitude ranges, at about 1200-1350 km and 1000-1100 km, the electron density measured during T118 were the lowest ever reported from Cassini at those altitude ranges, while the density measured during T119 seems rather normal compared to all the other profiles. The density altitude profiles are quite similar inbound and outbound, if looking at each flyby separately.

The middle and right panels of Figure \ref{figthree} shows the data split up into inbound and outbound legs, which illustrates that the observed difference in density is present at roughly the same altitude ranges both during the inbound leg, which occurred close to the day/night terminator of the moon, as well as during the outbound leg, which occurred further in toward the nightside of the moon. It is also interesting that at altitudes of around 1150 km during the inbound leg and about 1200 km during the outbound leg for T118, the electron density showed two distinct and similar peaks suggesting that the rise is not associated with a sporadic increase in the flux of precipitating magnetospheric particles. Two broader but much smaller-amplitude peaks were observed in the altitude range 1350-1600 km during both the inbound and the outbound leg. Also, at C/A, at an altitude of about 980 km, the densities seems to converge to similar values for the two flybys. For T119, at the same altitudes as were the peaks were located during T118, are now instead depletions of plasma. This is more pronounced during the outbound leg of T119. However interesting this coinciding dip appear, we should state that it could simply be a coincidence arising in an otherwise dynamic ionosphere.

\section{Discussion}
In summary, the observations on the ionospheric electron density during T118 and T119 show that (i) there is an overall large difference in density between the two flybys, (ii) the density difference is present both during the inbound legs and during the outbound legs, (iii) the density during T118 was unusually low compared to all other flybys, (iv) at certain altitude ranges was the electron density quite comparable between the two flybys, when distinct peaks appear in the T118 data. 
Next, we will discuss possible explanations for these observations of a large difference in electron density profiles and the distinct peaks during T118. 

Since most of the measurements presented in this paper were taken in Titan's nightside ionosphere, or at least at a SZA > 90 $^{\circ}$, the main ionisation source that is at play is particle impact ionisation, together with transport of plasma. The ambient conditions during T118 were classified as northern lobe (the same as during T119) but with some higher fluctuations in the magnetic field. These fluctuations could be indicative of Saturn's magnetospheric current sheet flapping past Titan during T118. The current sheet is typically filled with higher density plasma compared to the lobes, which could introduce additional particle impact ionisation in Titan's ionosphere. But since the density is lower during T118 this cannot be the explanation for the density differences. 

We can also note that the ion density derived from the LP (not shown) is generally also showing the same kind of profiles as the electron density, and the LP measurements of negative ions \citep{shebanits2013} do not show any distinct increases when the electron density decreases to abnormally low values during T118. This rules out any effect of dust, which could otherwise cause a depletion of free electrons as the charges attaches to the dust. The possibility exists that low mass negative ions are present as discussed by \citet{desai2017}, but as we cannot confirm that at this instance we leave that as an open question.

Even though these are nightside flybys, the SZA could still have some effect on the electron density such that when the SZA increases, the density should decrease. The SZA profile during T118 does go up to higher values (above 160$^{\circ}$) than during T119 on the outbound leg. However, during the inbound leg the SZA profiles are very similar and only differ by a few degrees. If looking at the flybys individually, the electron profiles are very similar on the inbound leg compared to the outbound leg, which suggests that the SZA is not important here. Also, since the electron density difference between the two flybys is present both during the inbound legs and the outbound legs, this implies that the SZA is not an important factor here.

Furthermore, the orientation of the ambient magnetic field is very similar during both flybys with roughly $\bf{B}$=[-2,-7,-2] nT (in TIIS) during the inbound leg, and similar values during the outbound leg (see Figure \ref{figtwo}d). Within Titan's ionosphere, the draping pattern is also similar for the two flybys, but with some differences in especially the $B_x$ component before C/A. A different draping pattern could mean that precipitating magnetospheric electrons or ions provide different ionisation patterns in Titan's ionosphere \citep{snowden2013b,snowden2014,regoli2016}, which would result in Cassini observing different densities along its path. However, how much of a difference in observed density along Cassini's path this would actually mean is difficult to estimate, and modeling this is also somewhat futile since the ambient magnetic conditions that are required as input in the models are actually very similar during the two passes. That the precipitation of magnetospheric electrons can vary significantly in Titan's ionosphere was demonstrated e.g., in \citet{vigren2016}, who showed examples when for similar ambient neutral conditions the electron impact ionization rate was comparable on the nightside and dayside, despite the lack of a photo-electron source on the nightside.

The fact that the electron density was unusually low during T118 would mean that either the ionisation rate was unusually low, that there was no transport of plasma to that region, that the recombination rate was increased, that significant ion outflow has occurred, that the neutral density was lower than normal, or some combination of the above. In case of photochemical equilibrium the density is expected to scale with the square root of the production rate divided by the effective recombination coefficient, so a reduction of $N_e$ by a dividing factor of 5 would require the ratio within the square root to be reduced by a dividing factor of 25. The electron temperature, as measured by LP are quite comparable between the flybys (down to the level of the accuracy of the instrument) and not able to cause a major difference in the recombination rate. Neutral density measurements by the INMS instrument (not shown) are complicated by a data gap during T119 from 1350 - 1900 km for some species. During the inbound leg (the outbound INMS data can be affected by wall effects in the instrument) the neutral density values are also quite comparable and the density during T118 is actually higher below 1100 km for especially $N_{2}$, so this does not seem like an explanation for the electron density difference between T118 and T119.  

The distinct peaks in density during T118 at 1150 km on the inbound leg and at 1200 km on  the outbound leg are also intriguing. They suggest a very localised plasma production regime, with low production both above and below, or transport to this specific altitude. We note again that the $B_x$ was different between the two flybys and could have an influence on the precipitation of impact ionising particles. Neutral density wave structures are observed in the INMS data during these flybys (not shown), which could also affect the plasma density. The vertical extent of the two distinct peaks in electron density during T118 are of typical wavelength for neutral waves \citep{muellerwodarg2006, cui2014}.

We stated above that the ambient Saturn magnetospheric conditions were similar during the two flybys, according to MAG data. However, one measured difference was that the energetic particle pressure was about 36$\%$ higher during the inbound leg of T118 compared to the inbound leg of T119 (0.88 pPa versus 0.56 pPa), as shown in Figure \ref{figtwo}g. This could possibly account for causing the larger gradient in the electron density observed in the topside ionosphere during T118, as the higher pressure compresses the ionospheric plasma. We note that the magnetic pressure and the thermal pressure are normally orders of magnitude larger than this. However, further down in the ionosphere, below the steep density gradient at about 1600-1700 km, the density during T118 is lower than during T119, which is somewhat counter-intuitive: if the ionospheric plasma is compressed then the density should be increased - but this does not seem to be the case here. However, it could still be that the higher pressure causes more dynamics to the ionosphere and affect the transport of plasma, and which would result in variable altitude profiles, as is seen during T118.

Finally, in this paper we have rather extensively discussed the difference between two particular flybys. But how does this pair of flybys compare to other pairs, or series, of flybys with similar geometry? We can briefly mention that T18-19, T41-T42, T64-T65 and T120-T121 were sets of passes with C/A near the terminator or on the dayside, and where differences in the density profiles were observed, while during T70-T71 and T55-T59 no significant or only smaller differences in electron density were observed in the ionosphere as a whole . 
Therefore, it is not uncommon to observe large differences in electron density altitude profiles as reported during T118 and T119, even though the flyby geometry is similar. What is striking with T118 and T119 is that the difference in density between the two passes is so large, and that no ambient conditions (that we can observe, anyway) seems to differ between the two instances.  

 \section{Conclusions}
There is a large difference in electron density between T118 and T119, even though the flyby geometry was similar. The difference is primarily due to an unusually low density during T118, but which cause is not easily explained from our limited data set. Since the flyby was a nightside pass we can suggest that suppressed ionisation from particle impacts (which is the main source of ionisation on the nightside), perhaps due to a less favorable magnetic field draping pattern, is a likely cause of the low density during T118 and the density difference between T118 and T119. This is not supported by any measurements, and it is in fact opposite to what the measurements of the high energy particles suggest, which showed a higher pressure during T118 compared to T119. Unfortunately, no measurements of lower energy particles (including electrons) were available for these flybys and we can only guess that they were significantly reduced at the time. We can also speculate on that the dynamics in the ionosphere is increased during T118 due to the higher ambient energetic plasma pressure and the presence of neutral density gravity waves. This could perturb the transport of plasma in the ionosphere and could be the cause of the two distinct peaks in electron density during T118.

\acknowledgments
The data used in this work is available on the NASA PDS. EV and DJA are grateful for funding from the Swedish National Space Board (Dnr 14/166 and 162/14). JC is supported by the National Science Foundation of China through grants 41525015 and 41774186. LHR contribution was supported by a NASA Living With a Star grant (NNX16AL12G). NJTE acknowledges support from the Swedish National Space Board (Dnr 135/13) and Vetenskapsr{\aa}det (621-2013-4191).






%
%
%
%
%
%
%
\bibliography{refs.bib} 
%
%
%





\listofchanges

\end{document}